\documentclass[twocolumn]{aastex63}

\newcommand\niproblem{{\rm Ni problem}}

\received{\today}

\submitjournal{ApJ}

\shorttitle{}
\shortauthors{Sawada and Suwa}

\usepackage{amsmath}
\usepackage{natbib}

\begin{document}

\title{A Consistent Modeling of Neutrino-driven Wind with Accretion Flow onto a Protoneutron Star and its Implications for $^{56}$Ni Production}

\correspondingauthor{Ryo Sawada}
\author{Ryo Sawada}
\affil{Department of Astrophysics and Atmospheric Sciences, Faculty of Science, Kyoto
Sangyo University, Motoyama, Kamigamo, Kita-ku, Kyoto 603-8555, Japan}
\email{ryo@cc.kyoto-su.ac.jp}
\author{Yudai Suwa}
\affil{Department of Astrophysics and Atmospheric Sciences, Faculty of Science, Kyoto
Sangyo University, Motoyama, Kamigamo, Kita-ku, Kyoto 603-8555, Japan}
\affil{Center for Gravitational Physics, Yukawa Institute for Theoretical Physics, Kyoto
University, Kitashirakawa Oiwakecho, Sakyo-ku, Kyoto 606-8502, Japan}

\begin{abstract}
Details of the explosion mechanism of core-collapse supernovae (CCSNe) are not yet fully understood. 
There is now an increasing number of successful examples of reproducing explosions in the first-principles simulations, which have shown a slow increase of explosion energy.
However, it was recently pointed out that the growth rates of the explosion energy of these simulations are insufficient to produce enough $^{56}$Ni mass to account for observations.
We refer to this issue as the `nickel mass problem' (\niproblem, hereafter) in this paper.
The neutrino-driven wind is suggested as one of the most promising candidates for the solution to the \niproblem\, in previous literature, but a multi-dimensional simulation for this is computationally too expensive to allow long-term investigations.
In this paper, we first built a consistent model of the neutrino-driven wind with an accretion flow onto a protoneutron star (PNS), by connecting a steady-state solution of the neutrino-driven wind and a phenomenological mass accretion model.
Comparing the results of our model with the results of first-principles simulations, we find that  the total ejectable amount of the neutrino-driven wind is roughly determined within $\sim$ 1 s from the onset of the explosion and the supplementable amount at a late phase ($t_e \gtrsim 1$ s) remains $M_\mathrm{ej} \lesssim 0.01M_\odot$ at most.
Our conclusion is that it is difficult to solve the \niproblem\, by continuous injection of $^{56}$Ni by the neutrino-driven wind.
We suggest that the total amount of synthesized $^{56}$Ni can be estimated robustly if simulations are followed up to $\sim 2$ s.
\end{abstract}
\keywords{supernovae: general}

\section{Introduction} \label{sec:intro}
Core-collapse supernovae (CCSNe) occur at the end of the lives of massive stars, 
lead to the birth of neutron stars and stellar black holes, and are the production sites of many elements. 
Details of the explosion mechanism of CCSNe, however, are not yet fully understood. 
The most promising scenario is the delayed neutrino driven explosion \citep{1985ApJ...295...14B}.
On the theoretical studies for the explosion mechanism, there is now an increasing number of successful examples of reproducing explosions in first-principles simulations. 
They solve multidimensional hydrodynamics equations, as well as a detailed neutrino transport  \citep[see, e.g., ][and references therein]{2012ARNPS..62..407J,2020arXiv200914157B}.
Most, if not all, of those state-of-the-art simulations, have shown a slow increase of explosion energy, and the growing rate of the explosion energy is typically $\mathcal{O}(0.1)$ Bethe $\mathrm{s}^{-1}$ ($1$ Bethe $=1\times 10^{51}$ erg), especially for 3D simulations.
On the other hand, one of the observational constraints is the amount of $^{56}$Ni, which drives supernova brightness.
A synthesized amount of $^{56}$Ni has been suggested to be very sensitive to the explosion properties and the progenitor core structure, since to synthesize $^{56}$Ni the temperature needs to be $T\gtrsim5\times10^9$ K
\citep[see e.g.][]{2012ApJ...757...69U,2016ApJ...818..124E,2016ApJ...821...38S,2019MNRAS.483.3607S}.
The amount of $^{56}$Ni has been measured from many SNe through light curves with reasonable accuracy \citep[see e.g.][]{2003ApJ...582..905H}. 
A typical amount of $^{56}$Ni obtained for well-studied SNe is on average $\sim0.07M_\odot$ \citep[e.g., SN 1987A, SN 1994I, SN 2002ap; ][]{1989ARA&A..27..629A,1994ApJ...437L.115I,2002ApJ...572L..61M}. 
Additionally, the statistical analysis for more than 50 events of CCSNe has also suggested that the amount of $^{56}$Ni is around $\sim0.07M_\odot$ at the median \citep{2019MNRAS.485.1559P}. 
It means that, in a canonical CCSNe, on average $\sim0.07M_\odot$ of $^{56}$Ni should be synthesized.
However, recent studies have shown that to reproduce the typical mass $0.07M_\odot$ of $^{56}$Ni by the explosive nucleosynthesis in the ejecta, the growth rate of the explosion energy of $\mathcal{O}(1)$ Bethe $\mathrm{s}^{-1}$ is required 
\citep{2019MNRAS.483.3607S,2019ApJ...886...47S}.  
In other words, the growth rate of the explosion energy of $\mathcal{O}(0.1)$ Bethe $\mathrm{s}^{-1}$, which is obtained in current typical explosion simulations, is insufficient to produce enough $^{56}$Ni mass.
We refer to this issue as the `nickel mass problem' (\niproblem, hereafter) in this paper.
While we should note that some models in first-principles simulations have succeeded in producing sufficient amounts of $^{56}$Ni \citep[e.g.,][]{2016ApJ...818..123B,2018JPhG...45a4001E,2020MNRAS.491.2715B,2020arXiv201010506B}, it is unclear whether we can reproduce sufficient $^{56}$Ni amount as a canonical nature.

The neutrino-driven wind, which is subject of this study, is thought to be one of the most promising candidates for the solution to the \niproblem\, \citep[e.g.,][]{2017ApJ...842...13W,2018ApJ...852...40W}.
The neutrino-driven wind is a phenomenon that blows out of the surface of the proto-neutron star (PNS) after the evacuation of the early ejecta in CCSNe.
This wind may solve the issue for the following two reasons.
First, the wind can provide $^{56}$Ni in addition to the hydrostatic and explosive nucleosynthesis, since it continues until $\sim$ 10 s after an explosion, which is longer than the converging time of the explosive nucleosynthesis ($\sim 1$ s).
Second, recent detailed simulations have predicted proton-rich ejecta in the post-explosion winds \citep[e.g.,][]{2016ApJ...818..123B} 
and also have indicated that almost all materials come to $^{56}$Ni in the wind with $Y_e\gtrsim  0.5$ \citep{2018ApJ...852...40W}.
But recent long-term spherical simulations of the PNS cooling phase show the rapidly decreasing neutrino luminosities and insufficient mass ejection \citep{2010A&A...517A..80F,2010PhRvL.104y1101H,2013ApJ...770L..22W} so that we may need multi-dimensional simulations, which take into account the mass accretion onto a PNS and ejection by the wind simultaneously, to solve this issue.
The problem of these multi-D simulations is that they have a computational time limitation and the possibility of the solution now lies in the phase later than the typical computational time.

In this paper, we investigate the potential of the neutrino-driven wind to solve \niproblem, especially at later phases than a few seconds after a successful explosion.
For this purpose,  we first build a consistent model of the neutrino-driven wind with the accretion flow onto the PNS, by connecting a steady-state solution of the neutrino-driven wind and a phenomenological mass accretion.
We then compare the results of our model with the results of first-principles simulations and discuss the possibilities to solve the \niproblem.
In Section \ref{sec:models}, we describe the treatment of three important equations in our modeling: the neutrino-driven winds, mass accretion flows onto the PNS, and combining them.
Our results are given in Section \ref{sec:results} before summary in Section \ref{sec:summary}.

\section{Models} \label{sec:models}
\begin{figure}[ht]
\begin{center}
  \includegraphics[width=80mm]{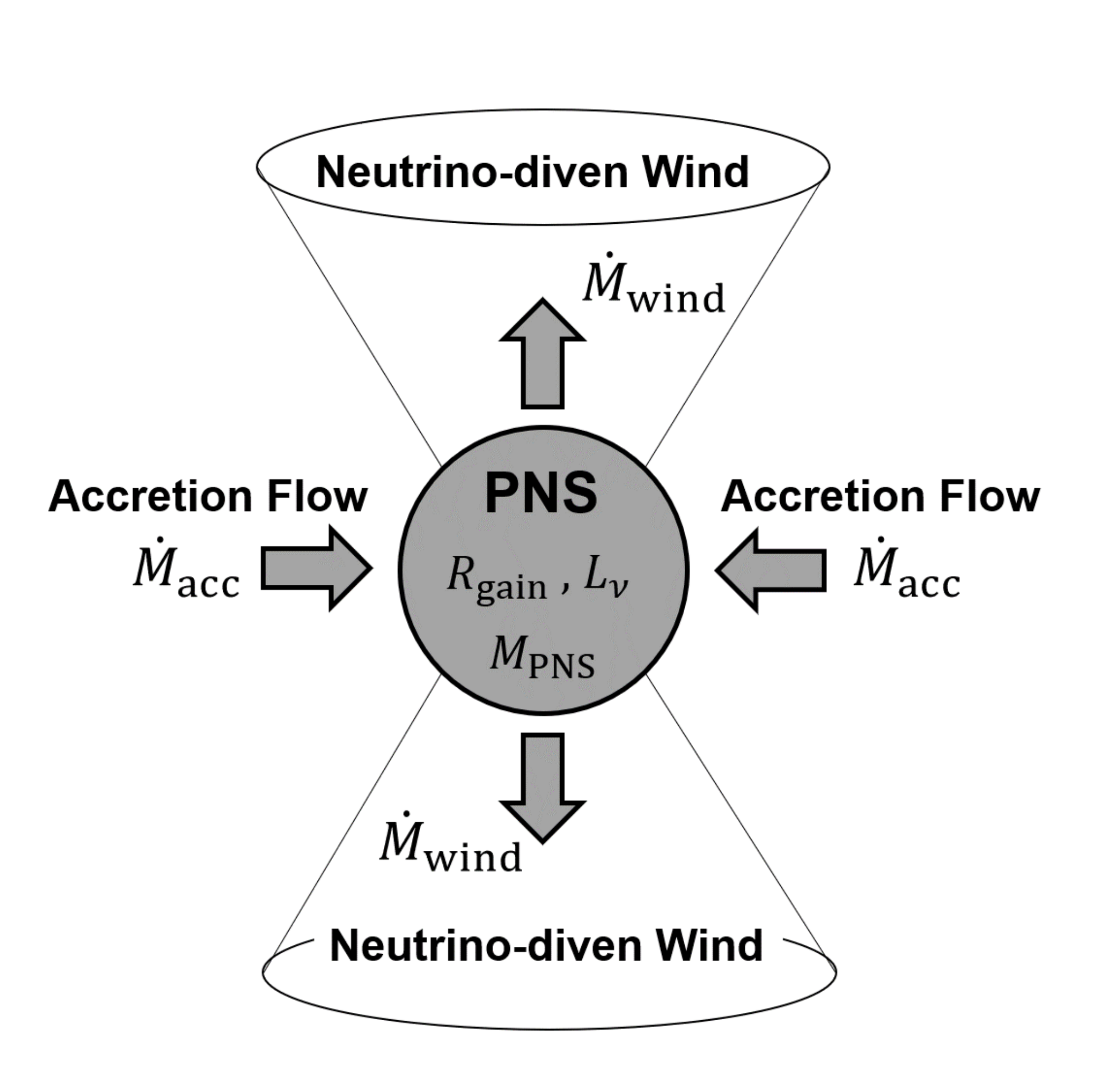}
    \caption{A schematic picture of a consistent model of the accretion flow onto the PNS and the neutrino-driven wind. 
    Our model assumes a system in which, the neutrino-driven winds blow out into the low-density region swept by the initial aspherical shock of CCSN explosion, and the mass accretion onto the PNS follows from the region where the initial shock did not develop.
    In order to build this model, we use the steady-state solution to describe the nature of neutrino-driven wind as a function of the PNS information $L_\nu$, $R_\mathrm{gain}$ and $M_\mathrm{PNS}$ (\S \ref{sec:wind}).
    We then formulate the phenomenological accretion flow model $\dot{M}_\mathrm{acc}$ (\S \ref{sec:acc}).
    Finally, we construct a consistent model by expressing the PNS information for the case with accretion flow as a function of mass accretion rates $\dot{M}_\mathrm{acc}$ and PNS masses $M_{\mathrm{PNS}}$, taking into account geometric effects (\S \ref{sec:PNS}).
}\label{fig:image} 
\end{center}
\end{figure}

In this section, we aim to build a consistent model of the neutrino-driven wind with the accretion flow onto the PNS, as illustrated in Fig. \ref{fig:image}.
Our model assumes a system in which, the neutrino-driven winds blow out into the low-density region swept by the initial aspherical shock of CCSN explosion, and the mass accretion onto the PNS follows from the region where the initial shock did not develop.
It is known that the neutrino-driven wind are very successfully described by the steady-state semi-analytical solutions \cite[e.g.,][]{1996ApJ...471..331Q,2001ApJ...554..578W,2001ApJ...562..887T,2013ApJ...770L..22W,2018ApJ...855..135B}, 
and now we have an interest in the nature of the neutrino-driven wind blowing radial direction into the low-pressure and low-density region swept by the SN shock.
Thus, the nature of the wind inside the blowing angle can be well-described by the spherically symmetric semi-analytic wind model.
To build this consistent model, 
we first solve the steady-state equations of the wind and derive relations between the wind $\dot{M}_\mathrm{wind}$ and the PNS profiles;
the neutrino luminosities $L_\nu$, 
the gain radius $R_\mathrm{gain}$ and 
the PNS mass $M_\mathrm{PNS}$ (\S \ref{sec:wind}).
The next step is to formulate the phenomenological accretion flow model $\dot{M}_\mathrm{acc}$ (\S \ref{sec:acc}).
We then model the evolution of the gain radius $R_\mathrm{gain}$ and the neutrino luminosity $L_\nu$ with accretion rates $\dot{M}_\mathrm{acc}$ and PNS masses $M_{\mathrm{PNS}}$, and finally derive the wind model with an accretion flow, taking into account geometric effects (\S \ref{sec:PNS}).

\subsection{Steady-state wind model} \label{sec:wind}
In this study, we use the spherically symmetric and general relativistic semi-analytic wind model in \cite{2013ApJ...770L..22W}.
We followed the \cite{2001ApJ...562..887T} for detailed calculation method.
Previous works studied the physical state of the neutrino wind for a wide range of neutron star masses and neutrino luminosities \citep[e.g.,][]{2000ApJ...533..424O,2018ApJ...855..135B}. 
Based on these results, they then studied the behavior of the neutrino-driven wind with the PNS evolution was studied by stitching this semi-analytic wind model \citep[e.g.,][]{2001ApJ...554..578W,2013ApJ...770L..22W}.

The basic equations to describe the spherically-symmetric and steady-state winds in the Schwarzschild geometry are
\begin{align}
    \dot{M}&=4\pi r^2 \rho v ~,\label{eq:mass}\\
    v\cfrac{dv}{dr}&=-\cfrac{1+(v/c)^2-(2GM/c^2r)}{\rho(1+\epsilon/c^2)+P/c^2}\cfrac{dP}{dr}-\cfrac{GM_\mathrm{PNS}}{r^2}~,\label{eq:moment}\\
    \dot{Q}&=v\left(\cfrac{d\epsilon}{dr}-\cfrac{P}{\rho^2}\cfrac{d\rho}{dr}\right)~\label{eq:energy},
\end{align}
where $\dot{M}$ is the constant mass outflow rate, 
$r$ is the distance from the center of the protoneutron star, 
$\dot{Q}$ is the heating rate, 
$\rho$ is the baryon mass density, 
$v$ is the radial velocity of the wind, 
$P$ is the pressure, and $\epsilon$ is the specific internal energy.
The system of Eqs. \eqref{eq:mass}–\eqref{eq:energy} is closed with the Helmholtz equation of state \citep{2000ApJS..126..501T}, which describes the stellar plasma as a mixture of arbitrarily degenerate and relativistic electrons and positrons, black-body radiation, and ideal Boltzmann gases of a defined set of fully ionized nuclei, taking into account corrections for the Coulomb effects.
The source term $\dot{Q}$ includes both heating and cooling by neutrino interactions.
Heating is due to the following three processes; i) neutrino and antineutrino captured by free nucleons, ii) neutrino scattering on electrons and positrons, and iii) neutrino-antineutrino pair annihilation into electron-positron pairs.
Cooling is due to electron and positron capture by free nucleons, and annihilation of electron-positron pairs into neutrino-antineutrino pairs \citep[for more details, see Eqs. (8)$-$(16) in][]{2000ApJ...533..424O}.

To determine the luminosity of each type of neutrino  $L_\nu$, we use the assumptions of $\dot{Y_e} = 0$ \citep{2018ApJ...855..135B}, 
in which we assume electron/positron captures are in equilibrium and an initial composition consists mainly of neutrons and protons.
Then, we get
\begin{equation}
    Y_e=\left[ 1+ \cfrac{L^n_{\overline{\nu}_e} \langle \sigma_{{\overline{\nu}_e} p}\rangle }{L^n_{\nu_e} \langle \sigma_{{\nu_e} n}\rangle}\right]^{-1} =0.5,\label{eq:Ye}
\end{equation}
where $L^n_\nu = L_\nu/ \langle E_\nu\rangle $ is the number luminosity and is assumed to be the same for electron neutrinos and anti-neutrinos. 
The cross-sections of electron neutrino absorption at neutrons ($\langle \sigma_{\nu_e n}\rangle$) and electron anti-neutrino absorption at protons ($\langle \sigma_{\overline{\nu}_e p}\rangle$) depend on the average neutrino and anti-neutrino energies.
Thus, given $L_{\nu_e}$, $\langle E_{\nu_e}\rangle$ and $Y_e$ as parameters, Eq. \eqref{eq:Ye} and the assumption of $L^n_{\nu_e}=L^n_{\overline{\nu}_e}$ leads to the anti-neutrino energy and luminosity.
With respect to the choice of $Y_e$, we are interested in an environment that maximizes the production of $^{56}$Ni.
Previous studies have shown that the abundance of $^{56}$Ni is extremely suppressed at $Y_e<0.5$ in nuclear statistical equilibrium (NSE) \citep{2008ApJ...685L.129S}, and the same behavior is known to occur in the neutrino-driven wind environment \citep{2018ApJ...852...40W}.
In order to focus on maximizing $^{56}$Ni production, we here fix $Y_e$ value to be $0.5$ throughout this paper.

Regarding the inner boundary condition, we assume, for simplicity, that the neutrino-driven wind starts at the gain radius $R_\mathrm{gain}$, which the heating and cooling due to neutrino interactions are in equilibrium ($\dot{Q}\approx0$). It is because the wind blows from the heating region outside the gain radius.
It allows us to determine the temperature of the inner boundary, $T_0$.
While the inner boundary is set to be $\rho_{0}=10^{10} \mathrm{g~cm^{-3}}$ in the ordinary neutrino-driven wind models \citep[e.g.,][]{2000ApJ...533..424O,2001ApJ...554..578W,2013ApJ...770L..22W},
we set the density at the inner boundary in our model to $\rho_{0}=10^{10} \mathrm{g~cm^{-3}} L_{\nu,51}^{1/2}$, following the method discussed by \cite{2015ApJ...810..115F}, 
in order to solve with the very high neutrino luminosity.
Given the density of the inner boundary $\rho_{0}$, Eq. \eqref{eq:mass} determines the initial velocity $v_0$ at $r=R_\mathrm{gain}$ for each $\dot{M}$.

The mass outflow rate $\dot{M}$ corresponds physically to determining how much material is ejected by the neutrino-driven wind, and algebraically to giving the behavior of the velocity in solving the equations.
In other words, the solutions of Eqs. \eqref{eq:mass}–\eqref{eq:energy} depend on this mass outflow rate $\dot{M}$ \citep{1996ApJ...471..331Q}.

\begin{figure}[ht]
\begin{center}
  \includegraphics[width=95mm]{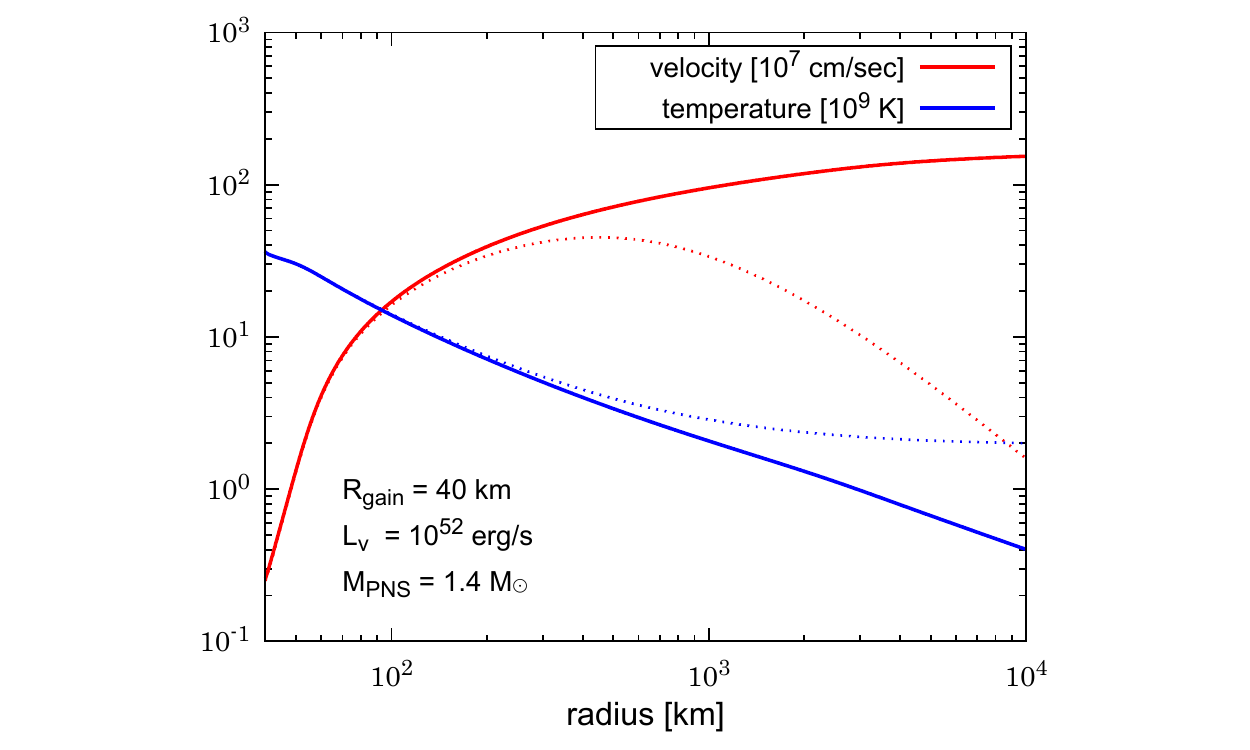}
\caption{
Outflow velocity $v(r)$ in units of $10^7$ cm s$^{-1}$ and temperature $T(r)$
in units of $10^9$ K as functions of the distance $r$ from the center of the PNS with the PNS mass $M_\mathrm{PNS}=1.4 M_\odot$, the gain radius $R_\mathrm{gain}=40$ km and the neutrino luminosity $L_\nu = 10^{52}$ ergs $\mathrm{s^{-1}}$.
Solid and dashed lines display the transonic and subsonic solutions, which corresponds to the mass outflow rate $\dot{M}_\mathrm{tran}= 8.25 \times 10^{-3}M_\odot~ \mathrm{s^{-1}}$ and $\dot{M}= 8.00 \times 10^{-3}M_\odot~ \mathrm{s^{-1}}$, respectively.
} \label{fig:NDW}
\end{center}
\end{figure}

For the case of small $\dot{M}$, the wind moves slowly, and $dv/dt$ can become zero before the velocity $v$ reaches the sonic speed $v_s$.
When this occurs, the velocity decreases after reaching a maximum value and becomes always subsonic (breeze or subsonic solution).
When $\dot{M}$ increases, it can happen that the acceleration term becomes zero at the same time as $v$ reaches $v_s$. 
This case corresponds to a critical value $\dot{M}=\dot{M}_\mathrm{tran}$ (transonic solutions). 
In this critical case, the velocity increases through the sound speed to supersonic values, eventually becoming a constant when almost the internal energy is converted into mechanical kinetic energy. 
Mass outflow rates larger than $\dot{M}_\mathrm{tran}$ are unphysical because, for these values of $\dot{M}$, $v$ has reached $v_s$ when the acceleration term is still positive, resulting in an infinite acceleration. 
Thus, we only need to focus our attention on cases of $\dot{M}\leqq\dot{M}_\mathrm{tran}$.
Here we briefly review two particular cases of the neutrino-driven wind with different mass outflow rates $\dot{M}$, which give transonic and subsonic solutions.
Figure \ref{fig:NDW} shows the fluid velocity and temperature as a function of the radius from the center of the PNS for each $\dot{M}$, where the PNS mass $M_\mathrm{PNS}=1.4 M_\odot$, the gain radius $R_\mathrm{gain}=40$ km and the neutrino luminosity $L_\nu = 10^{52}$ ergs $\mathrm{s^{-1}}$.
In this PNS parameters ($L_\nu$, $R_\mathrm{gain}$, $M_\mathrm{PNS}$), the transonic and subsonic solutions correspond to the mass outflow rate $\dot{M}=\dot{M}_\mathrm{tran}= 8.25 \times 10^{-3}M_\odot~ \mathrm{s^{-1}}$ and $\dot{M}= 8.00 \times 10^{-3}M_\odot~ \mathrm{s^{-1}}$ respectively.
Note that $\dot{M}=\dot{M}_\mathrm{tran}$ gives the maximum mass ejection.
In the following, we focus our discussion on transonic solutions which give the maximum ejected amount of the wind.

\begin{figure}[ht]
\begin{center}
  \includegraphics[width=95mm]{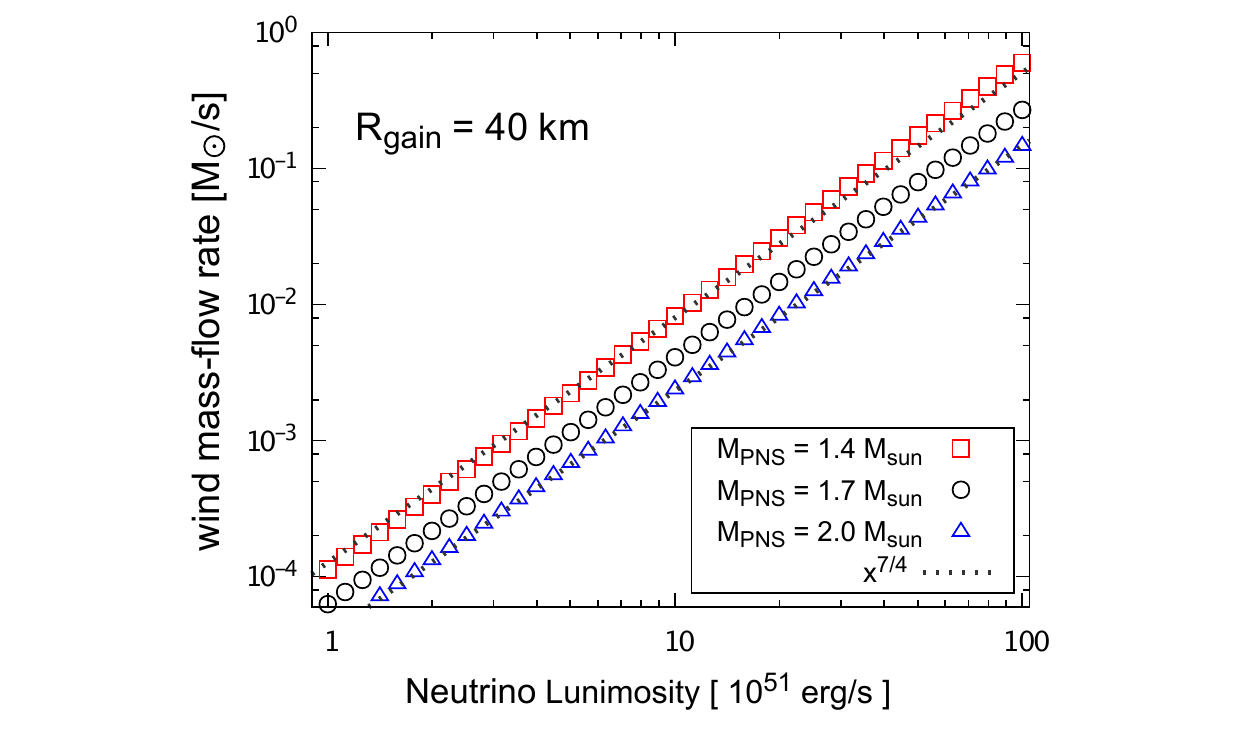}
  \includegraphics[width=95mm]{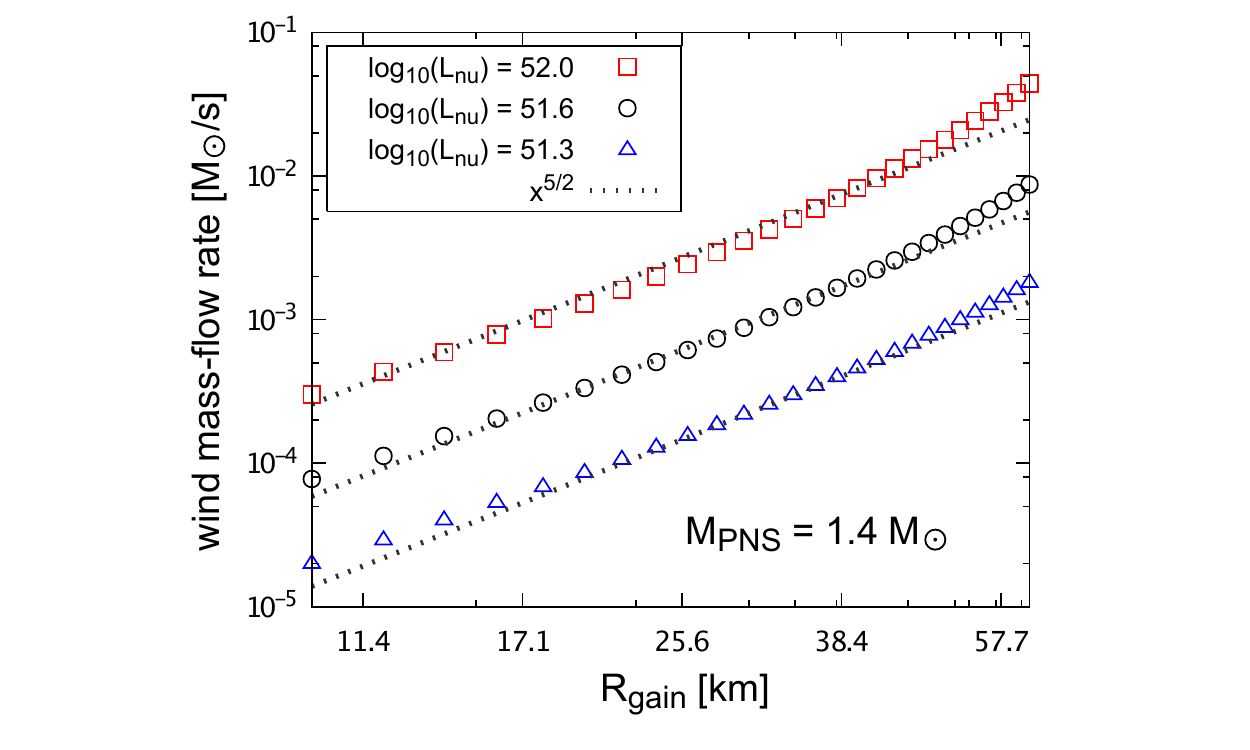}
  \includegraphics[width=95mm]{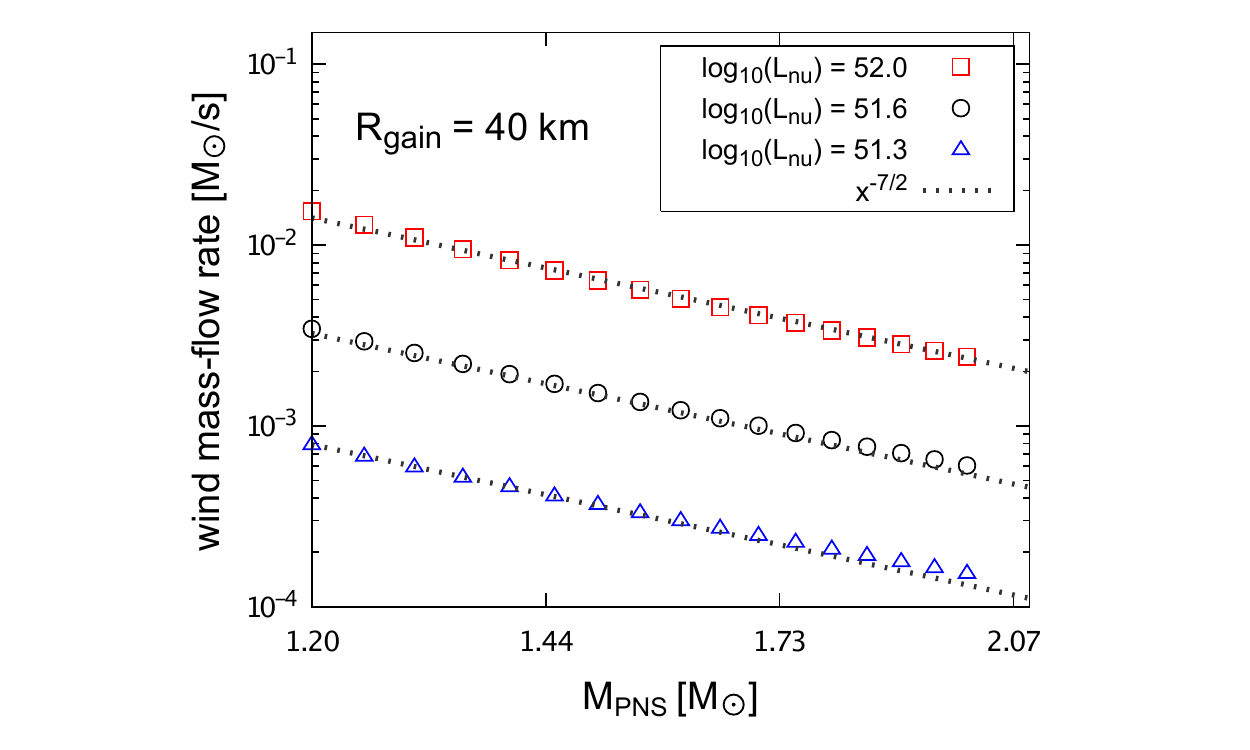}
\caption{Maximum mass flow rate $\dot{M}_\mathrm{wind,iso}(=\dot{M}_\mathrm{tran})$ as a function of $L_\nu$(top), $R_\mathrm{gain}$(middle), and $M_\mathrm{PNS}$(bottom) of the transonic wind, respectively.
Each type of point corresponds to the results of varying other parameters, which is different from the parameter chosen for the variable on the figure.
The dashed lines are the power-law relationship shown in equation \eqref{eq:result}, and can be seen to be roughly fitted to the plot.
} \label{fig:Mdot}
\end{center}
\end{figure}

The system of Eqs. \eqref{eq:mass}–\eqref{eq:energy} provides a (transonic) wind solution $\dot{M}_\mathrm{tran}$ for each set of three PNS parameters ($L_\nu$, $R_\mathrm{gain}$, $M_\mathrm{PNS}$).
In the following, we consider to model the relation between the maximum mass ejection of the wind $\dot{M}_\mathrm{wind,iso}=\dot{M}_\mathrm{tran}$ and the three PNS parameters.
Figures \ref{fig:Mdot} illustrates mass ejection rates $\dot{M}_\mathrm{wind,iso}$ as a function of each PNS parameters.
We confirmed that our results are in close agreement with the results of the previous 1D numerical \citep{2000PASJ...52..601S} 
and semi-analytical studies \citep{2000ApJ...533..424O,2001ApJ...554..578W} with the same set of parameters, except for the treatment of $Y_e$.

It is found that ejection rates $\dot{M}_\mathrm{wind,iso}$ have a relation with the PNS parameters, which is approximated by a power-law function as
\begin{align}
\dot{M}_{\mathrm{wind,iso}}
    &\approx 8.3 \times 10^{-3}M_\odot~ \mathrm{s^{-1}}\nonumber\\
    &\times \left(\cfrac{L_{\nu_e}}{10^{52}{\mathrm{erg~s^{-1}}}}\right)^\alpha
    \left(\cfrac{R_\mathrm{gain}}{4\times10^6\mathrm{cm}}\right)^\beta
    \left(\cfrac{M_\mathrm{PNS}}{1.4M_\odot}\right)^\gamma,
    \label{eq:result}
\end{align}
where the index we adopted is as follows; $\alpha=7/4$, $\beta=5/2$ and $\gamma=-7/2$.
To credit this modeling, we should mention two points in comparison to previous studies.
First, our model focuses on a mainly larger region for the gain radius $R_{\mathrm{gain}}$  ($10-60$ km), compared to previous studies \citep[$10-30$ km ;][]{1996ApJ...471..331Q,2000ApJ...533..424O}.
We confirmed that almost the same relation can be found as in \cite{1996ApJ...471..331Q} when we focus on the same parameter region of $R_{\mathrm{gain}}$ as the literature.
Second, while the result of \cite{2001ApJ...554..578W} tends to deviate the power-law relation at high luminosity, our solutions do not.
It is due to our treatment of the density at the inner boundary (gain radius) that follows the method of \cite{2015ApJ...810..115F}.
This method takes into account the physical equilibrium conditions, especially at high neutrino luminosity, which is different from those of the \cite{2001ApJ...554..578W}.
We also confirmed that the same trend as in \cite{2001ApJ...554..578W} when using the same boundary conditions.

Since we employ this power-law relation to construct our wind model in a later section, here we discuss the error between this equation and the numerical solution.
Table \ref{tbl:Mdot} shows the wind solution from the numerical calculations ($\dot{M}_{\mathrm{wind,cal}}$), the value estimated from the power-law relation ($\dot{M}_{\mathrm{wind,model}}$), and the error of $\dot{M}_{\mathrm{wind,cal}}$ with respect to $\dot{M}_{\mathrm{wind,model}}$, for typical values of each PNS parameters.
The error values in Table \ref{tbl:Mdot} indicate that our subsequent analytical discussion using the power-law relation keep an error within 30\% from the more precise numerical calculations.

\begin{table*}
\centering
\caption{The wind solution from the numerical calculations ($\dot{M}_{\mathrm{wind,cal}}$), the value estimated from the power-law relation ($\dot{M}_{\mathrm{wind,model}}$), and the error of $\dot{M}_{\mathrm{wind,cal}}$ with respect to $\dot{M}_{\mathrm{wind,model}}$, for typical values of each PNS parameters.}\label{tbl:Mdot}
  \begin{tabular}{rccccr} \hline\hline
    $L_{\nu_e}$&$R_\mathrm{gain}$&$M_\mathrm{PNS}$
    &$\dot{M}_{\mathrm{wind,cal}}$&$\dot{M}_{\mathrm{wind,model}}$&error $^a$\\
    $(10^{51}\mathrm{erg~s^{-1}})$&(km)&$(M_\odot)$
    &$(M_\odot{\mathrm{s^{-1}}})$&$(M_\odot{\mathrm{s^{-1}}})$&(\%)\\ \hline \hline
    10&40&1.4&$8.25 \times 10^{-3}$&$8.25 \times 10^{-3}$&0.0\\  \hline
    100&40&1.4&$5.98 \times 10^{-1}$&$4.64 \times 10^{-1}$&28.9\\ 
     1&40&1.4&$1.23 \times 10^{-4}$&$1.47 \times 10^{-4}$&-16.3\\  \hline
    10&50&1.4&$1.79 \times 10^{-2}$&$1.44 \times 10^{-3}$&24.4\\ 
    10&10&1.4&$3.00 \times 10^{-4}$&$2.58 \times 10^{-3}$&16.2\\  \hline
    10&40&2.0&$2.40 \times 10^{-3}$&$2.37 \times 10^{-3}$&1.3\\ 
    10&40&1.2&$1.54 \times 10^{-2}$&$1.42 \times 10^{-2}$&8.5\\ \hline\hline\\
  \end{tabular}
  	\tablecomments{
$^a$We denote the difference between the value of the numerical solution to the model value, normalized by the model value $\left(\frac{\dot{M}_{\mathrm{wind,cal}}-\dot{M}_{\mathrm{wind,model}}}{\dot{M}_{\mathrm{wind,model}}} \right)$, as a percentage.}
\end{table*}

\subsection{Mass accretion model}\label{sec:acc}
\begin{figure}[ht]
    \begin{center}
      \includegraphics[width=95mm]{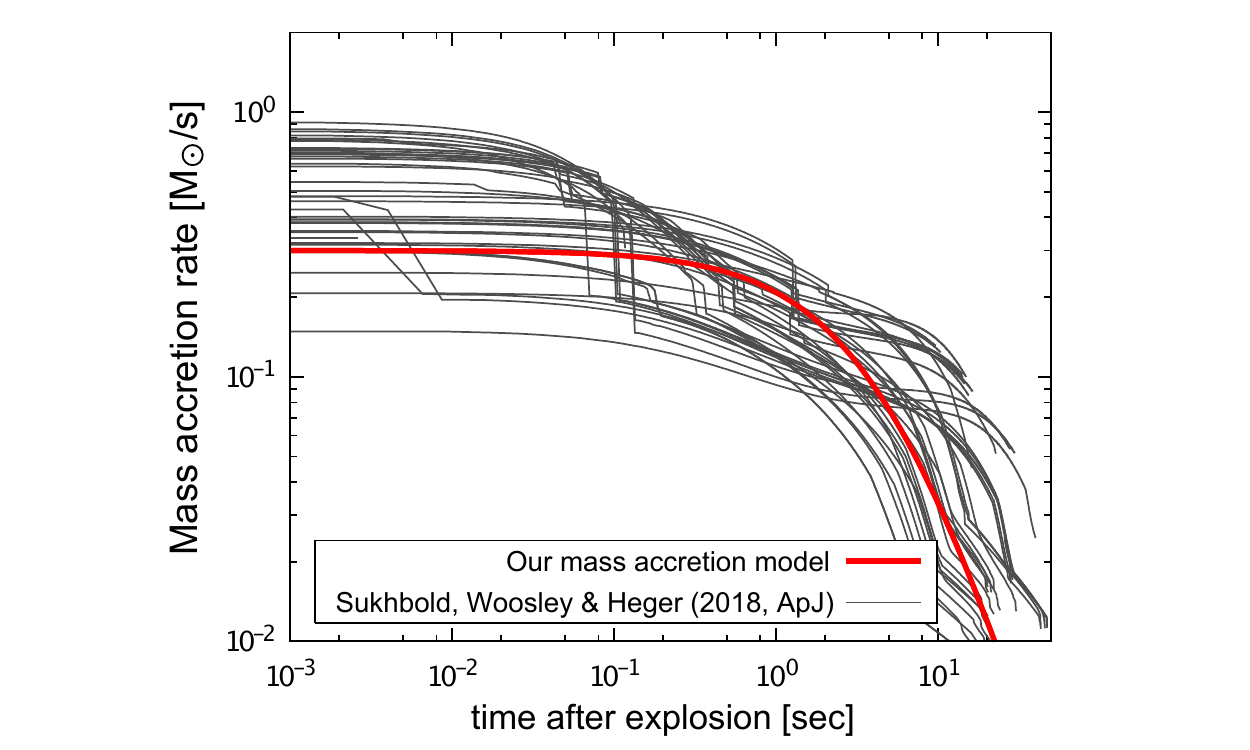}
    \caption{Mass accretion rate calculated by the free-falling model of \cite{2016MNRAS.460..742M}.
    The gray line is the mass accretion rate for the pre-SN stars of \cite{2018ApJ...860...93S} 
    (on $0.1M_\odot$ steps over the range $M_\mathrm{ZAMS}=12.0-20.0M_\odot$). 
    The red line is our simplified model of this mass accretion rate, and is given by $\dot{M}_{acc}(t)=\dot{M}_{acc,0}  \left(t/t_0+1\right)^{-2}$. 
} \label{fig:Macc}
    \end{center}
\end{figure}

In this section, we construct a phenomenological mass accretion model based on \cite{2016MNRAS.460..742M}.
We assume that matter reaches on the PNS with a free-fall timescale. 
The isotropic mass accretion rate $\dot{M}_\mathrm{acc,iso}$ is thus related to the mass coordinate $M$ of the infalling shell as \citep{2015ApJ...806..145W},
\begin{align}
\begin{split}
\dot{M}_{\mathrm{acc,iso}}
    &= \cfrac{2M}{t_f}~\cfrac{\rho}{\overline{\rho}-\rho}
    ~~\label{eq:acc},
\end{split}
\end{align}
where $\rho$ is the initial density of a given mass shell, 
$\overline{\rho}$ is the average density inside a given mass shell located at an initial radius $r$ (i.e. $M=4\pi \overline{\rho}r^3/3$ ), 
and $t_f$ is the infall time, which is defined as a function of the average density $\overline{\rho}$ inside a given mass shell by
\begin{align}
\begin{split}
t_f&= \sqrt{\cfrac{\pi}{4G\overline{\rho}}}~~.
\end{split}
\end{align}

Figure \ref{fig:Macc} shows the mass accretion rates $\dot{M}_{\mathrm{acc,iso}}$ as a function of $t$,
which is identified with $t_f$. 
In our model, the time origin is when the neutrino-driven wind starts to blow, corresponding to the time of the SN shock revival, which is given by the time at the mass shell of $M_{s=4k_B}$ accreting onto the PNS. 
Here $M_{s=4k_B}$ gives the mass coordinate with the entropy being $4 k_B$ baryon$^{-1}$.
This is because recent hydrodynamics simulations show that the shock launch takes place when a mass element with $s=4k_B$ baryon$^{-1}$ accretes onto the shock \citep{2016ApJ...818..124E,2016ApJ...816...43S}.
Gray lines are the mass accretion rates of progenitor stars from \cite{2018ApJ...860...93S} 
(on $0.1M_\odot$ steps over the range $M_\mathrm{ZAMS}=12.0-20.0M_\odot$).
We approximate our accretion model as the red line, which is given by
\begin{align}
    \dot{M}_{\mathrm{acc,iso}}(t)=
    \dot{M}_{\mathrm{acc,0}}\left(\cfrac{t}{t_0}+1\right)^{-2}
    ~~,\label{eq:accmodel}
\end{align}
where $\dot{M}_\mathrm{acc,0}$ and $t_0$ are free parameters.

\subsection{Consistent wind model with mass accretion} \label{sec:PNS}

Our description above is based on a one-dimensional radial flow. 
In order to construct a consistent model of the wind with mass accretion, it is necessary to take into account the geometric structure.
Hereafter, we present the collimation of the wind and the accretion flow due to the asymmetric structure of the supernova explosion with a geometrical factor $f_\Omega~(\leq 1)$. 
$f_\Omega$ relates the wind intrinsic properties and isotropic equivalents as
\begin{align}
    \dot{M}_{\mathrm{wind}}
    & = f_\Omega\dot{M}_{\mathrm{wind,iso}}~, \\
    \dot{M}_{\mathrm{acc}}
    & = (1-f_\Omega)\dot{M}_{\mathrm{acc,iso}} \label{eq:geo}~.
\end{align}

We next rewrite the outflow rate $\dot{M}_\mathrm{wind}$, which is obtained for a given set of the three parameters ($L_\nu$, $R_\mathrm{gain}$, $M_\mathrm{PNS}$) in Section \ref{sec:wind}, to the equation which is approximately determined by a given set of the two parameters ($\dot{M}_\mathrm{acc}$, $M_\mathrm{PNS}$).
We need the model of the gain radius $R_\mathrm{gain}$ and the neutrino luminosity $L_\nu$ as function of $\dot{M}_{\rm acc}$ and $M_\mathrm{PNS}$.
\cite{2016MNRAS.460..742M} found that, in the condition of $\dot{M}_{\rm acc}\gg 10^{-3}M_\odot \mathrm{s^{-1}}$, the gain radius $R_\mathrm{gain}$ can be described as
\begin{align}
    R_\mathrm{gain} 
    &\approx 40\, \mathrm{km} ~ 
    \left(\cfrac{\dot{M}_{\rm acc}}{0.1 M_\odot{\mathrm{s^{-1}}}}\right)^{1/3}
    \left(\cfrac{M_{\mathrm{PNS}}}{1.4M_\odot}\right)^{-1}~\label{eq:Rgain}.
\end{align}
This approximation has been confirmed to show reasonably consistent results with the contraction of the PNS in hydrodynamics simulations.

In our system, the neutrino luminosity $L_{\nu}$ is assumed to be dominated by accretion luminosity $L_{\nu,\mathrm{acc}}$.
The accretion luminosity $L_{\nu,\mathrm{acc}}$ is roughly given by the mass accretion rate $\dot{M}_{\rm acc}$ and the gravitational potential at the neutron star surface
\citep{2009A&A...499....1F},
\begin{equation}
    L_{\nu_e}\approx L_{\nu,\mathrm{acc}}
    =\eta \cfrac{GM_\mathrm{PNS}\dot{M}_{\rm acc}}{R_\mathrm{PNS}}~\label{eq:Lnu},
\end{equation}
where $R_\mathrm{PNS}=5R_\mathrm{gain}/7$ \citep{2016MNRAS.460..742M} and $\eta$ is an efficiency parameter, which specifies the conversion of accretion energy into neutrino luminosity. 
Note that the neutrino luminosity includes two components: accretion luminosity and diffusion luminosity \citep{2009A&A...499....1F}.
It is difficult to completely separate these components.
By taking into account the contribution of the diffusion luminosity $L_{\nu,\mathrm{diff}}$, $\eta$ would exceed unity  \citep[see ][]{2014ApJ...788...82M}.
In this study, we calibrate our model with the electron neutrino luminosity at $\sim1$ sec after the SN shock revival, 
which is well studied in the first-principles calculations of CCSN explosions \citep{2014ApJ...788...82M}, and use $\eta=1$.

From Eqs. \eqref{eq:result}--\eqref{eq:geo}, we can write the ejection rate of the neutrino-driven wind with the accretion flow onto the PNS as
\begin{align}
    \dot{M}_{\mathrm{wind}}
    &\approx 1.3\times10^{-2}~M_\odot\mathrm{s^{-1}} \nonumber\\
    &\times f_\Omega\left(\cfrac{(1-f_\Omega)\dot{M}_{\mathrm{acc,iso}}}{0.1 M_\odot{\mathrm{s^{-1}}}}\right)^{\frac{2\alpha+\beta}{3}}
    \left(\cfrac{M_{\mathrm{PNS}}}{1.4M_\odot}\right)^{2\alpha-\beta+\gamma},\label{eq:wind}
\end{align}
where the value of the indices estimated by adopting our model are $\frac{2\alpha+\beta}{3}=2$ and $2\alpha-\beta+\gamma=-5/2$.

\section{Total mass ejection by winds}\label{sec:results}
\subsection{Possible Contribution to the Ni problem}

In this section, we describe a potential of the neutrino-driven wind to solve the \niproblem.
A summary of the claim of the \niproblem~is that while on average $\sim0.07M_\odot$ of $^{56}$Ni should be synthesized in a canonical CCSNe, the first-principles simulations have been able to reproduce less than half of it.
Our purpose is to investigate whether the wind can eject $0.07M_\odot$ of $^{56}$Ni or not.

$^{56}$Ni is primarily synthesized where a material of an electron fraction $Y_e\approx0.5$ experiences NSE in a high temperature environment ($\gtrsim5\times10^9$ K). 
This is because in NSE it is dominated by Fe-peak nuclei due to the largest binding energy per nucleon and a proton-to-nucleon ratio being close to the electron fraction of the environment.
The neutrino-driven wind of interest now experiences NSE, as can be seen in Figure \ref{fig:NDW}, and is thus a sufficient environment for the synthesis of $^{56}$Ni.
In addition, detailed nucleosynthesis calculations confirmed the dominance of $^{56}$Ni in an outflow of electron fraction $Y_e\approx0.5$ \citep{2018ApJ...852...40W}.
Therefore, assuming that all the ejected material has $Y_e = 0.5$ as described in \S \ref{sec:wind}, it could be understand that the maximum ejected amount of $Y_e = 0.5$ material is available as a robust upper limit on the ejected $^{56}$Ni mass.
Namely, in this section, we estimate the ejectable maximum amount of $Y_e=0.5$ material (which corresponds to the maximum amount of $^{56}$Ni) by integrating the mass ejection rate of our wind model.

We integrate Eq. \eqref{eq:wind} with Eq. \eqref{eq:accmodel} from the shock revival time ($t =0$) as follows,
\begin{align}
    M_{\mathrm{ej,\infty}}
    &=\int_0^\infty dt~~\dot{M}_{\mathrm{wind}}(t)
    \nonumber\\
    &\approx 1.3\times10^{-2}~M_\odot\mathrm{s^{-1}} ~ f_\Omega
    \left(\cfrac{ (1-f_\Omega)\dot{M}_{\mathrm{acc,0}}}{0.1 M_\odot{\mathrm{s^{-1}}}}\right)^{2}
    \nonumber\\
    &~~\times\left(\cfrac{M_{\mathrm{PNS,0}}}{1.4M_\odot}\right)^{-5/2}\int_0^\infty dt
    \left(\cfrac{t}{t_0}+1\right)^{-4}\nonumber\\
    &= 4.3\times10^{-3}~M_\odot
    \nonumber\\
    &~~ \times f_\Omega
    \left(\dfrac{t_0 }{1~{\rm s}}\right)
    \left(\cfrac{(1-f_\Omega)\dot{M}_{\mathrm{acc,0}}}{0.1 M_\odot{\mathrm{s^{-1}}}}\right)^{2}
    \left(\cfrac{M_{\mathrm{PNS,0}}}{1.4M_\odot}\right)^{-5/2}~,\label{eq:intg1}
\end{align}
where we neglect the mass evolution of the PNS within the integration, which decreases the mass ejection by $\mathcal{O}(10)\%$.
Moreover, when we adopt $f_\Omega=1/3$, which gives the geometric effect term $f_\Omega(1-f_\Omega)^2$ maximum, Eq. \eqref{eq:intg1} is written as 
\begin{align}
    M_{\mathrm{ej,\infty}}^{\mathrm{max}} 
    &= 6.4\times10^{-4}~M_\odot 
    \nonumber\\
    &~~\times \left(\dfrac{t_0 }{1~{\rm s}}\right) \left(\cfrac{\dot{M}_{\mathrm{acc,0}}}{0.1 M_\odot{\mathrm{s^{-1}}}}\right)^{2}
    \left(\cfrac{M_{\mathrm{PNS,0}}}{1.4M_\odot}\right)^{-5/2}.\label{eq:windmax1}
\end{align}

The main goal of this paper is to find out the ejectable maximum mass of our wind model from Eq. \eqref{eq:windmax1}.
When we adopt the initial mass of the PNS to be $1.4 M_\odot$ \citep[e.g.,][]{2014ApJ...788...82M}, then two free parameters, $\dot{M}_{\mathrm{acc,0}}$ and $t_0$, remain.
We first adopt $\dot{M}_{\mathrm{acc,0}}=1M_\odot~\mathrm{s^{-1}}$ for  $\dot{M}_{\mathrm{acc,0}}$ as a phenomenological upper limit from Fig. \ref{fig:Macc}.
$t_0$ is constrained by the maximum PNS mass as follows.
Taking into account the geometric effects (Eq. \ref{eq:geo}), and ignoring the mass decreases due to ejected wind because it is relatively small, the mass of the PNS can be written as
\begin{align}
    M_\mathrm{PNS}(t)
    &\approx M_\mathrm{PNS,0} + \int_0^t dt'~~\dot{M}_{\mathrm{acc}}(t') \nonumber\\
    &= M_\mathrm{PNS,0} + (1-f_\Omega)\dot{M}_{\mathrm{acc,0}} t_0\left(\cfrac{t}{t+t_0}\right)
    ,\label{eq:PNSmodel}
\end{align}
where $M_\mathrm{PNS,0}$ is the initial mass of the PNS (at the time of the SN shock revival).
We assume that the wind ceases when the PNS mass exceeds a black hole mass.
These assumptions of the total accretion mass (Eq. \ref{eq:PNSmodel} and the PNS mass up to $2.1M_\odot$) gives us $(1-f_\Omega)\dot{M}_{\mathrm{acc,0}}\,t_0\leq 0.7M_\odot$.
To conclude, the ejectable maximum mass of our wind model is given with $\dot M_{\rm acc,0}=1M_\odot\,{\rm s}^{-1}$, $t_0=1.05$ s, and $f_\Omega=1/3$ in Eq. \eqref{eq:windmax1} as follows,
\begin{align}
    M_{\mathrm{ej,\infty}}^{\mathrm{max}} 
    &= 0.067~M_\odot~~.\label{eq:windmax2}
\end{align}
If most of this compensation from the wind is added at late phase, which is later than the computation time of the first-principles simulations, this value is then sufficient to compensate for the lack of $^{56}$Ni in the recent \niproblem.
In the following, we investigate the time evolution of the cumulative ejected mass of the wind and discuss the nature of the explosion that could solve \niproblem.

\begin{figure}[ht]
    \begin{center}
      \includegraphics[width=95mm]{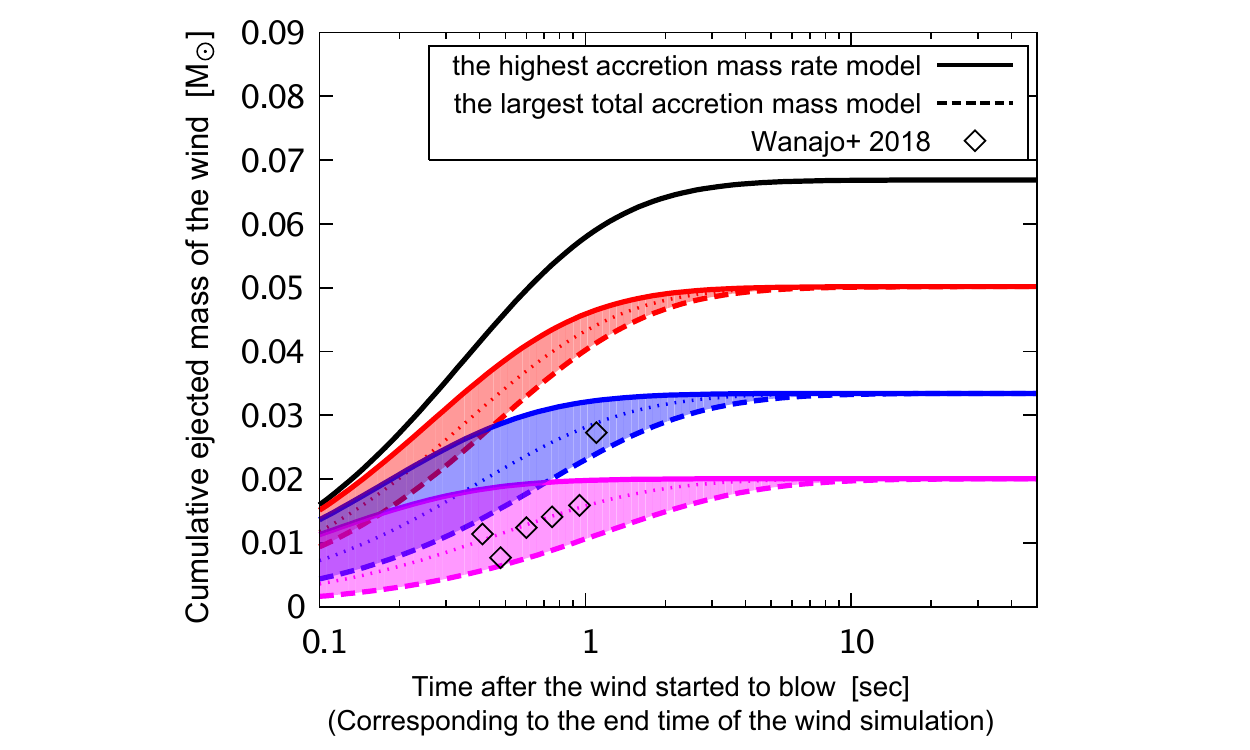}
    \caption{Cumulative ejected mass of the wind as a function of time after the neutrino-driven wind starts to blow (corresponding to the time after the SN shock revival).
    The parameter sets with the same total ejected mass is illustrated in the same color.
    In each color, the solid line represents the model with the largest mass accretion rate ($\dot{M}_{\mathrm{acc,0}}= 1.0M_\odot\mathrm{s^{-1}}$), 
    the dashed line represents the model with the largest total accretion mass ($(1-f_\Omega)\dot{M}_{\mathrm{acc,0}}t_0= 0.7M_\odot$), 
    and the dotted line corresponds to the model with the intermediate-total accretion mass.
    For instance, in the case of blue, the solid line: $\dot{M}_{\mathrm{acc,0}}= 1.0M_\odot\mathrm{s^{-1}}$ and $(1-f_\Omega)\dot{M}_{\mathrm{acc,0}}t_0= 0.35M_\odot$, 
    the dashed line: 
    $\dot{M}_{\mathrm{acc,0}}= 0.5M_\odot\mathrm{s^{-1}}$ and $(1-f_\Omega)\dot{M}_{\mathrm{acc,0}}t_0= 0.7M_\odot$, 
    and the dotted line: 
    $\dot{M}_{\mathrm{acc,0}}= 0.666M_\odot\mathrm{s^{-1}}$ and $(1-f_\Omega)\dot{M}_{\mathrm{acc,0}}t_0= 0.525M_\odot$.
    } \label{fig:Mwind}
    \end{center}
\end{figure}

Figure \ref{fig:Mwind} shows the time evolution of the cumulative ejected mass of the wind model.
The cumulative ejected mass is given as 
\begin{align}
    M_{\mathrm{ej}}(t_e)
    &=\int_0^{t_e} dt~~\dot{M}_{\mathrm{wind}}(t)
    \nonumber\\
    &\approx 6.4\times10^{-4} M_\odot
    \left[1-\left(\cfrac{t_0}{t_0+t_e}\right)^3\right]
    \nonumber\\
    &~~ \times \left(\frac{t_0}{1\,{\rm s}}\right)
    \left(\cfrac{\dot{M}_{\mathrm{acc,0}}}{0.1 M_\odot{\mathrm{s^{-1}}}}\right)^{2}
    \left(\cfrac{M_{\mathrm{PNS,0}}}{1.4M_\odot}\right)^{-5/2}~,\label{eq:intg2}
\end{align}
where $t_e$ is time after the neutrino-driven wind starts to blow, corresponding to the time after the SN shock revival. We adopt $f_\Omega=1/3$, which gives the geometric effect term $f_\Omega(1-f_\Omega)^2$ maximum.
When we fix $M_\mathrm{PNS,0} = 1.4M_\odot$ for simplicity, then two free parameters, $\dot{M}_{\mathrm{acc,0}}$ and $t_0$, remain to determine the trajectory of the time evolution.
As with the condition for Eq. \eqref{eq:windmax2}, these two parameters are given by the conditions 
$\dot{M}_{\mathrm{acc,0}}\leq 1.0M_\odot\mathrm{s^{-1}}$ and $(1-f_\Omega)\dot{M}_{\mathrm{acc,0}}\,t_0\leq 0.7M_\odot$ , respectively, from the phenomenological accretion model (Fig. \ref{fig:Macc}) and the limits of the total accretion mass (Eq. \ref{eq:PNSmodel} and the PNS mass up to $2.1M_\odot$).
A degenerate set of parameters that converge to the same $M_{\mathrm{ej,\infty}}$ are shown in the same color in Figure \ref{fig:Mwind}.
We further compare the time evolution with the multi-D first-principles simulations especially at $t_e \lesssim 1$ s \citep{2018ApJ...852...40W}, which are shown in Figure \ref{fig:Mwind} as rhombus points.
It indicates that, within parameter ambiguities, the total ejectable amount of the neutrino-driven wind is roughly determined within 1 s from the onset of the blowing, which is reachable for first-principles simulations.
Moreover, we also find that the supplementable amount from the wind at a later phase ($t_e \gtrsim 1$ s) remains $\lesssim 0.01M_\odot$.
It also shows that, comparing with first-principles simulations, the expected total amount from the wind in a canonical CCSN explosion is about $\lesssim 0.03 M_\odot$ at most.

We should mention that, in fact, some models in first-principles simulations of CCSNe have produced a sufficient amount of $^{56}$Ni in total of explosive and the wind nucleosynthesis  
\citep{2016ApJ...818..123B,2018JPhG...45a4001E,2020MNRAS.491.2715B,2020arXiv201010506B}.
However, the claim of the \niproblem~is that we are now struggling to reproduce sufficient $^{56}$Ni amount as a canonical nature of the CCSNe explosion, and many previous studies place their hopes on the supplement from the neutrino-driven wind at a later phase \citep[e.g.,][]{2017ApJ...842...13W,2018ApJ...852...40W}.
Our conclusion on this issue is that, in order to compensate for sufficient $^{56}$Ni by the neutrino-driven wind, it is preferred to have an active ejection in the early phase rather than a continuous ejection until the later phase.
We note that while the wind driven by magneto-rotational explosions may be more energetic, it is not expected to solve the \niproblem~due to low $^{56}$Ni as it is ejected with a low electron fraction \citep{2015ApJ...810..109N}.
It should also be emphasized that the total amount of synthesized $^{56}$Ni can be estimated robustly if the first-principles simulations are performed up to $\sim2$ s.

\subsection{Effects of PNS mass evolution}
To further expand the discussion in Eqs. \eqref{eq:intg1} and \eqref{eq:intg2} about the ejectable amount of the wind, 
here we discuss the effect of the time evolution of the PNS mass on the results above.
We first introduce the ratio $\epsilon$ of the total accretion mass to the initial PNS mass as
\begin{align}
    \epsilon
    =\cfrac{1}{M_\mathrm{PNS,0}} \int_0^\infty dt ~~    
    \dot{M}_{\mathrm{acc}}(t)
    ~~.\label{eq:fatrate}
\end{align}
Using this ratio $\epsilon$, the time evolution of the PNS mass expressed in Eq. \eqref{eq:PNSmodel} is written as
\begin{align}
    M_\mathrm{PNS}(t)
    &\approx M_\mathrm{PNS,0}\left(1 + \epsilon~\cfrac{t}{t+t_0}\right)
    ~~,\label{eq:PNSmodel2}
\end{align} 
where we note that this ignores the mass loss due to wind, as in Eq. \eqref{eq:PNSmodel}.
No matter how light the initial PNS mass is assumed at SN shock revival (e.g.,  $M_\mathrm{PNS,0}\sim1.2M_\odot$), 
this $\epsilon$ is less than unity, since this mass will never grow more than twice as large due to the constraint that the wind stops when it becomes a black hole.
Then we can update the time evolution term of the integral in Eq. \eqref{eq:intg1} as follows
\begin{align}
    \int_0^\infty dt~~&
    \left(\cfrac{t}{t_0}+1\right)^{-4}\left(1 + \epsilon~\cfrac{t}{t+t_0}\right)^{-5/2}    \nonumber\\
    &=\cfrac{2\left( \epsilon^2-4\epsilon+8\sqrt{\epsilon+1}-8 \right)}{3\epsilon^3} t_0 ~\nonumber\\
    &\approx \left( \cfrac{1}{3} -\cfrac{5}{24}\epsilon +\cfrac{7}{48}\epsilon^2  \right) t_0~~.
    \label{eq:epsilon}
\end{align}
where $\epsilon=0$ corresponds to the result in Eq. \eqref{eq:intg1}. We find that, for $\epsilon<1$, the effect of mass dependence only decreases the wind mass loss.
The same can be argued for the time evolution of the cumulative mass, i.e., the discussion in Eq. \eqref{eq:intg2} and Fig. \ref{fig:Mwind}.

\begin{figure}[htbp]
    \begin{center}
      \includegraphics[width=95mm]{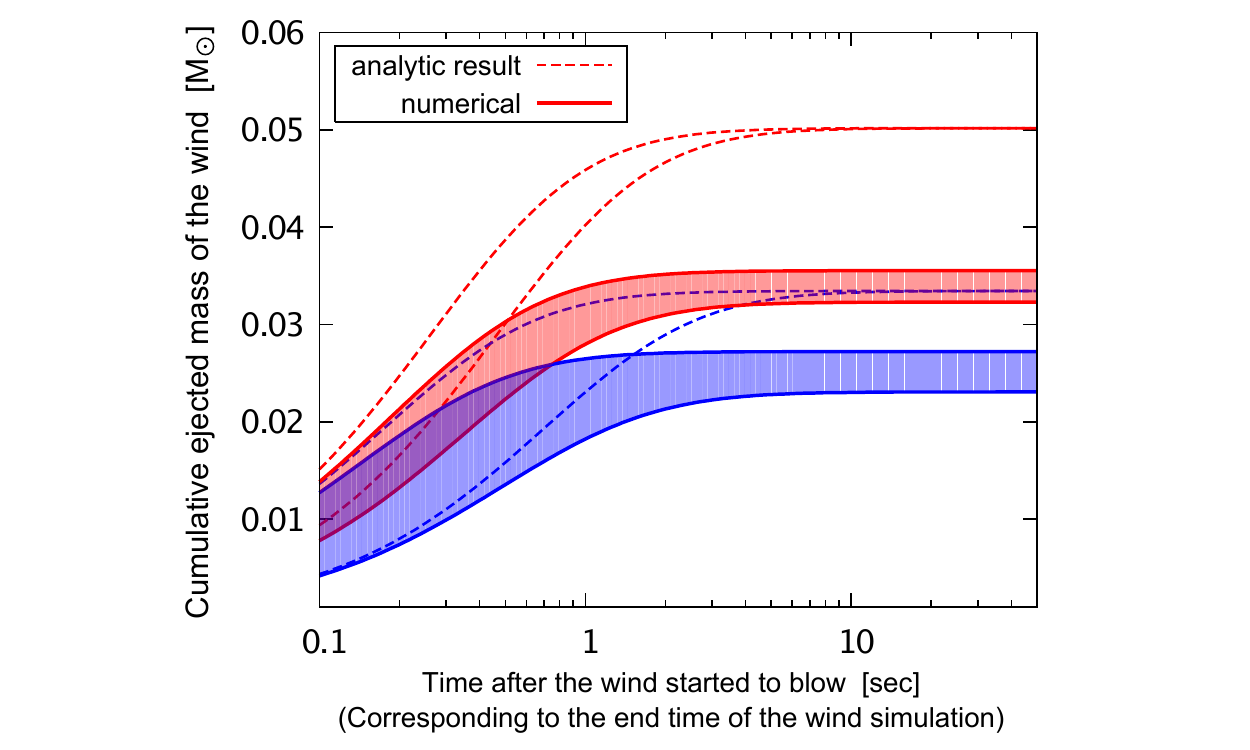}
    \caption{Same as Figure \ref{fig:Mwind}, but a comparison of the analytical results ignoring the effect of the time-evolution of PNS mass (Eq. \ref{eq:intg2}) and the numerical results taking into account the effect (see Eq. \ref{eq:epsilon}).} \label{fig:Mwind-num}
    \end{center}
\end{figure}

Fig. \ref{fig:Mwind-num} shows a comparison of the results between analytical integration without the effect of the PNS masses and the numerical integration with the effect.
This figure shows that the result of Eq. \eqref{eq:intg1} gives a robust upper limit on the ejectable mass of the wind.
This comparison indicates that the results of Eqs. \eqref{eq:intg1} and \eqref{eq:intg2} give robust upper limits on the ejectable mass of the wind.

\section{Discussion and Summary}\label{sec:summary}
In this paper, we have constructed a consistent model of the accretion flow onto the PNS and the neutrino-driven wind, and then estimate the ejectable $^{56}$Ni mass as a supplement to the \niproblem\ of the CCSN explosions.
We first derived the ejecta amount for the spherical-symmetric steady-state neutrino-driven wind with the PNS parameters as follow
\begin{align}
&\dot{M}_{\mathrm{wind,iso}}
    \approx ~ 
    8.3 \times 10^{-3}M_\odot \mathrm{s^{-1}}~\nonumber\\
    &\times \left(\cfrac{L_{\nu_e}}{10^{52}{\mathrm{erg~s^{-1}}}}\right)^{7/4}
    \left(\cfrac{R_\mathrm{gain}}{4\times10^6\mathrm{cm}}\right)^{5/2}
    \left(\cfrac{M_\mathrm{PNS}}{1.4M_\odot}\right)^{-7/2}
    ~~.\label{eq:sum1}
\end{align}
We then derived a consistent model of the neutrino-driven wind with the accretion flow onto the PNS by modeling the evolution of the gain radius $R_\mathrm{gain}$ and the neutrino luminosity $L_{\nu_e}$.
The wind mass loss rate (Figure \ref{fig:image}) is
\begin{align}
    \dot{M}_{\mathrm{wind}}\approx ~
    & 
    1.3\times10^{-2}~M_\odot\mathrm{s^{-1}} ~\nonumber\\
    &~~ \times f_\Omega
    \left(\cfrac{(1-f_\Omega)\dot{M}_{\mathrm{acc,iso}}}{0.1 M_\odot{\mathrm{s^{-1}}}}\right)^{2}
    \left(\cfrac{M_{\mathrm{PNS,0}}}{1.4M_\odot}\right)^{-5/2}~~.\label{eq:sum2}
\end{align}
Based on this equation and adopt $f_\Omega=1/3$, which gives the geometric effect term $f_\Omega(1-f_\Omega)^2$ maximum, the total ejectable amount is estimated as follows
\begin{align}
    M_{\mathrm{ej,\infty}}
    = &
    6.4\times10^{-4}~M_\odot  ~\nonumber\\
    &~~\times \left(\frac{t_0}{1\,{\rm s}}\right)  
    \left(\cfrac{\dot{M}_{\mathrm{acc,0}}}{0.1 M_\odot{\mathrm{s^{-1}}}}\right)^{2}
    \left(\cfrac{M_{\mathrm{PNS,0}}}{1.4M_\odot}\right)^{-5/2}~,\label{eq:sum3}
\end{align}
where $\dot{M}_{acc,0}$ and $t_0$ are the parameters which characterize the mass accretion rate $\dot{M}_{\mathrm{acc,iso}}$.
Eqs. \eqref{eq:sum1}--\eqref{eq:sum3} are important results in a consistent model of the neutrino-driven wind with the accretion flow onto the PNS (Fig. \ref{fig:image}), which is one of the goals of this paper.

Based on these equations, when we take the upper limit as far as possible within the range of the parameter constraint $M_{\mathrm{PNS,0}} \geq1.4M_\odot$, $\dot{M}_{\mathrm{acc,0}}\leq 1.0M_\odot\mathrm{s^{-1}}$ and $(1-f_\Omega)\dot{M}_{\mathrm{acc,0}} t_0\leq 0.7M_\odot$, the ejectable maximum mass of our wind model is derived to be $M_{\mathrm{ej,\infty}}= 0.067~M_\odot$.
If most of this compensation from the wind is added at late phase, which is later than the computation time of the first-principles calculations, this value is then sufficient to compensate for the lack of $^{56}$Ni in the recent \niproblem. 
However, we found that, within parameter ambiguities, the total ejectable amount of the neutrino-driven wind is roughly determined within 1 s from the start of the blowing, which is reachable by first-principles simulations.
Moreover, we also found that the supplementable amount from the wind at a later phase ($t_e \gtrsim 1$ s) remains $M_\mathrm{ej} \lesssim 0.01M_\odot$ at most, independent of the parameter choice, i.e., the nature of the explosion.
Our conclusions on the \niproblem~are first that in order to compensate for sufficient $^{56}$Ni by the neutrino-driven wind, it is preferred to have an active ejection in the early phase rather than a continuous ejection until the later phase.
It is also an important suggestion that the total amount of synthesized $^{56}$Ni can be estimated robustly if the first-principles simulations are followed up to 2 s.

Lastly, we mention the recent work by \cite{2020arXiv201010506B}, which was submitted after this paper was submitted. 
In their simulations, they observed the downflow and outflow system instead of spherical neutrino-driven wind and the outflow supplied $\lesssim 0.05 M_\odot$ of $^{56}$Ni. 
Our aspherical wind model fails to represent their downflow-outflow model, since our boundary condition assumes that the wind is slowly moving outward and is gradually accelerated by the neutrino heating in the vicinity of the gain radius. On the other hand, their downflow-ouflow system would produce smooth transition from the incoming flow to the outgoing flow and these mass flows shows almost the same value (see Fig. 1 of \citealt{2020arXiv201010506B}). 
It corresponds to $\dot M_{\rm wind}=\dot M_{\rm acc}$ in our model, which cannot be reproduced under the current set of equations with the efficiency neutrino parameter, $\eta$. 
In the forthcoming paper, we will present a model which takes into account such a situation.

In summary, it is difficult to solve the \niproblem~in a way that continuously supplements for $^{56}$Ni at the late phase of the explosion by the neutrino-driven wind.
Therefore, requiring an explosion capable of active ejection in the early phase of shock revival is a simple and straightforward solution for the SN mechanism to satisfy the \niproblem\ as canonical, without fine-tuning.

\acknowledgments
The work has been supported by Japan Society for the Promotion of Science (JSPS) KAKENHI Grant 19J14179 (R.S.), 18H05437, 20H00174, 20H01904 and 20H04747 (Y.S.).

\bibliography{RS2020}

\begin{thebibliography}{}
\expandafter\ifx\csname natexlab\endcsname\relax\def\natexlab#1{#1}\fi
\providecommand{\url}[1]{\href{#1}{#1}}
\providecommand{\dodoi}[1]{doi:~\href{http://doi.org/#1}{\nolinkurl{#1}}}
\providecommand{\doeprint}[1]{\href{http://ascl.net/#1}{\nolinkurl{http://ascl.net/#1}}}
\providecommand{\doarXiv}[1]{\href{https://arxiv.org/abs/#1}{\nolinkurl{https://arxiv.org/abs/#1}}}

\bibitem[{{Arnett} {et~al.}(1989){Arnett}, {Bahcall}, {Kirshner}, \&
  {Woosley}}]{1989ARA&A..27..629A}
{Arnett}, W.~D., {Bahcall}, J.~N., {Kirshner}, R.~P., \& {Woosley}, S.~E. 1989,
  \araa, 27, 629, \dodoi{10.1146/annurev.aa.27.090189.003213}

\bibitem[{{Bethe} \& {Wilson}(1985)}]{1985ApJ...295...14B}
{Bethe}, H.~A., \& {Wilson}, J.~R. 1985, \apj, 295, 14, \dodoi{10.1086/163343}

\bibitem[{{Bliss} {et~al.}(2018){Bliss}, {Witt}, {Arcones}, {Montes}, \&
  {Pereira}}]{2018ApJ...855..135B}
{Bliss}, J., {Witt}, M., {Arcones}, A., {Montes}, F., \& {Pereira}, J. 2018,
  \apj, 855, 135, \dodoi{10.3847/1538-4357/aaadbe}

\bibitem[{{Bollig} {et~al.}(2020){Bollig}, {Yadav}, {Kresse}, {Janka},
  {Mueller}, \& {Heger}}]{2020arXiv201010506B}
{Bollig}, R., {Yadav}, N., {Kresse}, D., {et~al.} 2020, arXiv e-prints,
  arXiv:2010.10506.
\newblock \doarXiv{2010.10506}

\bibitem[{{Bruenn} {et~al.}(2016){Bruenn}, {Lentz}, {Hix}, {Mezzacappa},
  {Harris}, {Messer}, {Endeve}, {Blondin}, {Chertkow}, {Lingerfelt},
  {Marronetti}, \& {Yakunin}}]{2016ApJ...818..123B}
{Bruenn}, S.~W., {Lentz}, E.~J., {Hix}, W.~R., {et~al.} 2016, \apj, 818, 123,
  \dodoi{10.3847/0004-637X/818/2/123}

\bibitem[{{Burrows} {et~al.}(2020){Burrows}, {Radice}, {Vartanyan}, {Nagakura},
  {Skinner}, \& {Dolence}}]{2020MNRAS.491.2715B}
{Burrows}, A., {Radice}, D., {Vartanyan}, D., {et~al.} 2020, \mnras, 491, 2715,
  \dodoi{10.1093/mnras/stz3223}

\bibitem[{{Burrows} \& {Vartanyan}(2020)}]{2020arXiv200914157B}
{Burrows}, A., \& {Vartanyan}, D. 2020, arXiv e-prints, arXiv:2009.14157.
\newblock \doarXiv{2009.14157}

\bibitem[{{Eichler} {et~al.}(2018){Eichler}, {Nakamura}, {Takiwaki}, {Kuroda},
  {Kotake}, {Hempel}, {Cabez{\'o}n}, {Liebend{\"o}rfer}, \&
  {Thielemann}}]{2018JPhG...45a4001E}
{Eichler}, M., {Nakamura}, K., {Takiwaki}, T., {et~al.} 2018, Journal of
  Physics G Nuclear Physics, 45, 014001, \dodoi{10.1088/1361-6471/aa8891}

\bibitem[{{Ertl} {et~al.}(2016){Ertl}, {Janka}, {Woosley}, {Sukhbold}, \&
  {Ugliano}}]{2016ApJ...818..124E}
{Ertl}, T., {Janka}, H.~T., {Woosley}, S.~E., {Sukhbold}, T., \& {Ugliano}, M.
  2016, \apj, 818, 124, \dodoi{10.3847/0004-637X/818/2/124}

\bibitem[{{Fischer} {et~al.}(2009){Fischer}, {Whitehouse}, {Mezzacappa},
  {Thielemann}, \& {Liebend{\"o}rfer}}]{2009A&A...499....1F}
{Fischer}, T., {Whitehouse}, S.~C., {Mezzacappa}, A., {Thielemann}, F.~K., \&
  {Liebend{\"o}rfer}, M. 2009, \aap, 499, 1,
  \dodoi{10.1051/0004-6361/200811055}

\bibitem[{{Fischer} {et~al.}(2010){Fischer}, {Whitehouse}, {Mezzacappa},
  {Thielemann}, \& {Liebend{\"o}rfer}}]{2010A&A...517A..80F}
---. 2010, \aap, 517, A80, \dodoi{10.1051/0004-6361/200913106}

\bibitem[{{Fujibayashi} {et~al.}(2015){Fujibayashi}, {Yoshida}, \&
  {Sekiguchi}}]{2015ApJ...810..115F}
{Fujibayashi}, S., {Yoshida}, T., \& {Sekiguchi}, Y. 2015, \apj, 810, 115,
  \dodoi{10.1088/0004-637X/810/2/115}

\bibitem[{{Hamuy}(2003)}]{2003ApJ...582..905H}
{Hamuy}, M. 2003, \apj, 582, 905, \dodoi{10.1086/344689}

\bibitem[{{H{\"u}depohl} {et~al.}(2010){H{\"u}depohl}, {M{\"u}ller}, {Janka},
  {Marek}, \& {Raffelt}}]{2010PhRvL.104y1101H}
{H{\"u}depohl}, L., {M{\"u}ller}, B., {Janka}, H.~T., {Marek}, A., \&
  {Raffelt}, G.~G. 2010, \prl, 104, 251101,
  \dodoi{10.1103/PhysRevLett.104.251101}

\bibitem[{{Iwamoto} {et~al.}(1994){Iwamoto}, {Nomoto}, {H{\"o}flich},
  {Yamaoka}, {Kumagai}, \& {Shigeyama}}]{1994ApJ...437L.115I}
{Iwamoto}, K., {Nomoto}, K., {H{\"o}flich}, P., {et~al.} 1994, \apjl, 437,
  L115, \dodoi{10.1086/187696}

\bibitem[{{Janka}(2012)}]{2012ARNPS..62..407J}
{Janka}, H.-T. 2012, Annual Review of Nuclear and Particle Science, 62, 407,
  \dodoi{10.1146/annurev-nucl-102711-094901}

\bibitem[{{Mazzali} {et~al.}(2002){Mazzali}, {Deng}, {Maeda}, {Nomoto},
  {Umeda}, {Hatano}, {Iwamoto}, {Yoshii}, {Kobayashi}, {Minezaki}, {Doi},
  {Enya}, {Tomita}, {Smartt}, {Kinugasa}, {Kawakita}, {Ayani}, {Kawabata},
  {Yamaoka}, {Qiu}, {Motohara}, {Gerardy}, {Fesen}, {Kawabata}, {Iye},
  {Kashikawa}, {Kosugi}, {Ohyama}, {Takada-Hidai}, {Zhao}, {Chornock},
  {Filippenko}, {Benetti}, \& {Turatto}}]{2002ApJ...572L..61M}
{Mazzali}, P.~A., {Deng}, J., {Maeda}, K., {et~al.} 2002, \apjl, 572, L61,
  \dodoi{10.1086/341504}

\bibitem[{{M{\"u}ller} {et~al.}(2016){M{\"u}ller}, {Heger}, {Liptai}, \&
  {Cameron}}]{2016MNRAS.460..742M}
{M{\"u}ller}, B., {Heger}, A., {Liptai}, D., \& {Cameron}, J.~B. 2016, \mnras,
  460, 742, \dodoi{10.1093/mnras/stw1083}

\bibitem[{{M{\"u}ller} \& {Janka}(2014)}]{2014ApJ...788...82M}
{M{\"u}ller}, B., \& {Janka}, H.-T. 2014, \apj, 788, 82,
  \dodoi{10.1088/0004-637X/788/1/82}

\bibitem[{{Nishimura} {et~al.}(2015){Nishimura}, {Takiwaki}, \&
  {Thielemann}}]{2015ApJ...810..109N}
{Nishimura}, N., {Takiwaki}, T., \& {Thielemann}, F.-K. 2015, \apj, 810, 109,
  \dodoi{10.1088/0004-637X/810/2/109}

\bibitem[{{Otsuki} {et~al.}(2000){Otsuki}, {Tagoshi}, {Kajino}, \&
  {Wanajo}}]{2000ApJ...533..424O}
{Otsuki}, K., {Tagoshi}, H., {Kajino}, T., \& {Wanajo}, S.-y. 2000, \apj, 533,
  424, \dodoi{10.1086/308632}

\bibitem[{{Prentice} {et~al.}(2019){Prentice}, {Ashall}, {James}, {Short},
  {Mazzali}, {Bersier}, {Crowther}, {Barbarino}, {Chen}, {Copperwheat},
  {Darnley}, {Denneau}, {Elias-Rosa}, {Fraser}, {Galbany}, {Gal-Yam},
  {Harmanen}, {Howell}, {Hosseinzadeh}, {Inserra}, {Kankare}, {Karamehmetoglu},
  {Lamb}, {Limongi}, {Maguire}, {McCully}, {Olivares E}, {Piascik}, {Pignata},
  {Reichart}, {Rest}, {Reynolds}, {Rodr{\'{\i}}guez}, {Saario}, {Schulze},
  {Smartt}, {Smith}, {Sollerman}, {Stalder}, {Sullivan}, {Taddia}, {Valenti},
  {Vergani}, {Williams}, \& {Young}}]{2019MNRAS.485.1559P}
{Prentice}, S.~J., {Ashall}, C., {James}, P.~A., {et~al.} 2019, \mnras, 485,
  1559, \dodoi{10.1093/mnras/sty3399}

\bibitem[{{Qian} \& {Woosley}(1996)}]{1996ApJ...471..331Q}
{Qian}, Y.~Z., \& {Woosley}, S.~E. 1996, \apj, 471, 331, \dodoi{10.1086/177973}

\bibitem[{{Sawada} \& {Maeda}(2019)}]{2019ApJ...886...47S}
{Sawada}, R., \& {Maeda}, K. 2019, \apj, 886, 47,
  \dodoi{10.3847/1538-4357/ab4da3}

\bibitem[{{Seitenzahl} {et~al.}(2008){Seitenzahl}, {Timmes},
  {Marin-Lafl{\`e}che}, {Brown}, {Magkotsios}, \&
  {Truran}}]{2008ApJ...685L.129S}
{Seitenzahl}, I.~R., {Timmes}, F.~X., {Marin-Lafl{\`e}che}, A., {et~al.} 2008,
  \apjl, 685, L129, \dodoi{10.1086/592501}

\bibitem[{{Sukhbold} {et~al.}(2016){Sukhbold}, {Ertl}, {Woosley}, {Brown}, \&
  {Janka}}]{2016ApJ...821...38S}
{Sukhbold}, T., {Ertl}, T., {Woosley}, S.~E., {Brown}, J.~M., \& {Janka}, H.-T.
  2016, \apj, 821, 38, \dodoi{10.3847/0004-637X/821/1/38}

\bibitem[{{Sukhbold} {et~al.}(2018){Sukhbold}, {Woosley}, \&
  {Heger}}]{2018ApJ...860...93S}
{Sukhbold}, T., {Woosley}, S.~E., \& {Heger}, A. 2018, \apj, 860, 93,
  \dodoi{10.3847/1538-4357/aac2da}

\bibitem[{{Sumiyoshi} {et~al.}(2000){Sumiyoshi}, {Suzuki}, {Otsuki},
  {Terasawa}, \& {Yamada}}]{2000PASJ...52..601S}
{Sumiyoshi}, K., {Suzuki}, H., {Otsuki}, K., {Terasawa}, M., \& {Yamada}, S.
  2000, \pasj, 52, 601, \dodoi{10.1093/pasj/52.4.601}

\bibitem[{{Suwa} {et~al.}(2019){Suwa}, {Tominaga}, \&
  {Maeda}}]{2019MNRAS.483.3607S}
{Suwa}, Y., {Tominaga}, N., \& {Maeda}, K. 2019, \mnras, 483, 3607,
  \dodoi{10.1093/mnras/sty3309}

\bibitem[{{Suwa} {et~al.}(2016){Suwa}, {Yamada}, {Takiwaki}, \&
  {Kotake}}]{2016ApJ...816...43S}
{Suwa}, Y., {Yamada}, S., {Takiwaki}, T., \& {Kotake}, K. 2016, \apj, 816, 43,
  \dodoi{10.3847/0004-637X/816/1/43}

\bibitem[{{Thompson} {et~al.}(2001){Thompson}, {Burrows}, \&
  {Meyer}}]{2001ApJ...562..887T}
{Thompson}, T.~A., {Burrows}, A., \& {Meyer}, B.~S. 2001, \apj, 562, 887,
  \dodoi{10.1086/323861}

\bibitem[{{Timmes} \& {Swesty}(2000)}]{2000ApJS..126..501T}
{Timmes}, F.~X., \& {Swesty}, F.~D. 2000, \apjs, 126, 501,
  \dodoi{10.1086/313304}

\bibitem[{{Ugliano} {et~al.}(2012){Ugliano}, {Janka}, {Marek}, \&
  {Arcones}}]{2012ApJ...757...69U}
{Ugliano}, M., {Janka}, H.-T., {Marek}, A., \& {Arcones}, A. 2012, \apj, 757,
  69, \dodoi{10.1088/0004-637X/757/1/69}

\bibitem[{{Wanajo}(2013)}]{2013ApJ...770L..22W}
{Wanajo}, S. 2013, \apjl, 770, L22, \dodoi{10.1088/2041-8205/770/2/L22}

\bibitem[{{Wanajo} {et~al.}(2001){Wanajo}, {Kajino}, {Mathews}, \&
  {Otsuki}}]{2001ApJ...554..578W}
{Wanajo}, S., {Kajino}, T., {Mathews}, G.~J., \& {Otsuki}, K. 2001, \apj, 554,
  578, \dodoi{10.1086/321339}

\bibitem[{{Wanajo} {et~al.}(2018){Wanajo}, {M{\"u}ller}, {Janka}, \&
  {Heger}}]{2018ApJ...852...40W}
{Wanajo}, S., {M{\"u}ller}, B., {Janka}, H.-T., \& {Heger}, A. 2018, \apj, 852,
  40, \dodoi{10.3847/1538-4357/aa9d97}

\bibitem[{{Wongwathanarat} {et~al.}(2017){Wongwathanarat}, {Janka},
  {M{\"u}ller}, {Pllumbi}, \& {Wanajo}}]{2017ApJ...842...13W}
{Wongwathanarat}, A., {Janka}, H.-T., {M{\"u}ller}, E., {Pllumbi}, E., \&
  {Wanajo}, S. 2017, \apj, 842, 13, \dodoi{10.3847/1538-4357/aa72de}

\bibitem[{{Woosley} \& {Heger}(2015)}]{2015ApJ...806..145W}
{Woosley}, S.~E., \& {Heger}, A. 2015, \apj, 806, 145,
  \dodoi{10.1088/0004-637X/806/1/145}

\end{thebibliography}
\bibliographystyle{aasjournal}

\end{document}